\documentclass[12pt]{article}
\usepackage{amsmath,amssymb,amsfonts,graphicx,bm}
\usepackage{pgfplots}
\usepackage{graphicx,bm}
\usepackage{geometry}

    \newgeometry{vmargin={20mm,30mm}, hmargin={25mm,25mm}}  
    
    \usepackage[margin=1cm]{caption}

\usepackage{float}
\usepackage{epsfig}
\usepackage{esint}
\usepackage{color}

\usepackage{multirow}

  \usepackage{paralist}
  \usepackage{epstopdf}
  \usepackage{graphics} 
 \usepackage[colorlinks=true]{hyperref}
 \hypersetup{urlcolor=blue, citecolor=red}
 \usepackage[latin1]{inputenc}
\usepackage[caption=false]{subfig}

 \usepackage{tikz-cd}

\usepackage{float}
\usepackage{epsfig}
\usepackage{esint}

\usepackage{bm,braket}

\newcommand{\R}{{\mathbb R}}

\newcommand{\e}{{\rm e}}

\newcommand{\K}{{\rm K}}

\newcommand{\calN}{{\mathcal N}}
\newcommand{\calM}{{\mathcal M}}

\newcommand{\calV}{{\mathcal V}}

\newcommand{\x}{\mathbf{x}}
\newcommand{\y}{\mathbf{y}}

\renewcommand{\P}{\mathbb{P}}

\newcommand{\E}{{\mathbb E}}

\newcommand{\calZ}{\mathcal Z}
\newcommand{\calT}{\mathcal T}

\begin{document}
\title{Kuramoto model with stochastic resetting and coupling through an external medium}

\author{ \em
Paul. C. Bressloff, \\ Department of Mathematics, Imperial College London, \\
London SW7 2AZ, UK}

\maketitle

\begin{abstract}

Most studies of collective phenomena in oscillator networks focus on directly coupled systems as exemplified by the classical Kuramoto model. However, there are growing number of examples in which oscillators interact indirectly via a common external medium, including bacterial quorum sensing (QS) networks, pedestrians walking on a bridge, and centrally coupled lasers. In this paper we analyze the effects of stochastic phase resetting on a Kuramoto model with indirect coupling. All the phases are simultaneously reset to their initial values at a random sequence of times generated from a Poisson process. On the other hand, the external environmental state is not reset. We first derive a continuity equation for the population density in the presence of resetting and show how the resulting density equation is itself subject to stochastic resetting. We then use an Ott-Antonsen (OA) ansatz to reduce the infinite-dimensional system to a four-dimensional piecewise deterministic system with subsystem resetting. The latter is used to explore how synchronization depends on a cell density parameter. (In bacterial QS this represents the ratio of the population cell volume and the extracellular volume.)  At high densities we recover the OA dynamics of the classical Kuramoto model with global resetting. On the other hand, at low densities, we show how subsystem resetting has a major effect on collective synchronization, ranging from noise-induced transitions to slow/fast dynamics.
 \end{abstract}

\section{Introduction}

\begin{figure}[t!]
\centering
\includegraphics[width=12cm]{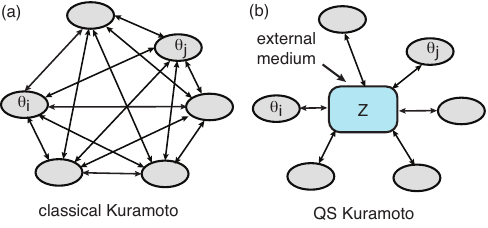} 
\caption{Schematic illustration of the Kuramoto model with (a) direct pairwise coupling and (b) indirect coupling via an external medium with complex amplitude $Z$. The latter is an example of a Quorum sensing (QS) network.}
\label{fig1}
\end{figure}

A number of natural and synthetic systems involve populations of oscillators that interact indirectly via a common external medium rather than via direct connections. One well-known example is a quorum sensing (QS) network of genetically engineered bacteria \cite{Ojalvo04,Hasty10} that communicate via signaling molecules (autoinducers) diffusing in the extracellular environment. Other examples include pedestrians walking on a bridge \cite{Strogatz05,Strogatz07}, which was inspired by the Millennium Bridge Problem in London, semiconductor lasers coupled through a central hub \cite{Roy10}, and social networks under the influence of social media \cite{Xue20}. Most theoretical studies of collective phenomena in oscillator networks focus on directly coupled systems \cite{Pikovsky03}, although there are some notable exceptions \cite{Russo10,Schwab12,Schwab12a,Sharma16,Verma19}. In particular, Schwab et al. \cite{Schwab12a} consider a modified version of the classical Kuramoto model \cite{Kuramoto84,Strogatz00,Acebron05}, in which the oscillators couple to an external medium that is represented by a complex amplitude $Z(t)=R(t)\e^{i\Phi(t)}$, see Fig. \ref{fig1}. As in the original Kuramoto model, each oscillator is described by a phase $\theta_j(t)$ and natural frequency $\omega_j$, $j=1,\dots,\calN$. Moreover, the natural frequencies are drawn from a prescribed probability density $h(\omega)$ with mean $\omega_0$. Following previous studies of direct coupling, the authors investigate the existence of partially coherent states in the continuum limit $\calN\rightarrow \infty$. The resulting continuity equation for the density of oscillators is solved by projecting the dynamics onto the corresponding Ott-Antonsen (OA) manifold with $h(\omega)$ taken to be Lorentzian. In contrast to the effective one-dimensional OA manifold found for direct coupling\cite{Ott08}, the OA manifold for indirect coupling is three-dimensional with coordinates $(r(t),R(t),\psi(t))$ where $\psi(t)=\Phi(t)-\phi(t)$ and $r(t)\e^{i\phi(t)}$ is the first circular moment of the phases. Consequently,  certain phenomena arise that are not found in the original model\cite{Schwab12a}, including bistability.

A key parameter in the Kuramoto model with indirect coupling is an effective cell density $\sigma_0$ that acts as a multiplier of the coupling constant in the dynamics of $Z(t)$. The origin of $\sigma_0$ can be understood by considering the application to bacterial quorum sensing, where $\sigma_0$ represents the ratio of the population cell volume to the extracellular volume. (Note that $\sigma_0$ is dimensionless and is denoted by $\rho$ in Ref. \cite{Schwab12a}.) In the high-density limit $\sigma_0\rightarrow \infty$ one finds that the OA dynamics reduces to the 1D dynamics of the standard Kuramoto model. On the other hand, in the low density regime, the system exhibits bistability between an incoherent state and a highly coherent state, or between coherent states with different amplitudes $R$. Moreover, in the limit $\sigma_0\rightarrow 0^+$ the model can be mapped to a Kuramoto model with a bimodal frequency distribution \cite{Martens09}. Finally, if the collective oscillation frequency of the oscillators is the same as the oscillation frequency $\Omega$ of the external medium ($\omega_0=\Omega$), then the model can be related to the Millennium Bridge Problem \cite{Strogatz05,Strogatz07}.

In this paper we extend the analysis of the Kuramoto model with coupling through an external medium to include the effects of stochastic phase resetting. That is, the phases $\theta_j(t)$ simultaneously reset to their initial values $\theta_{j,0}$ at a random sequence of times generated from a Poisson process with constant rate $\lambda$. We distinguish global phase resetting from a local resetting protocol in which each oscillator resets independently according to its own sequence
of resetting times. Note that there is growing interest in understanding the effects of global or local resetting on interacting particle systems (see the recent review \cite{Nagar23}), which builds upon the original work on single particle resetting \cite{Evans11a,Evans11b,Evans14,Evans20}. Several recent papers have analyzed the standard Kuramoto model with stochastic resetting \cite{Sarkar22,Bressloff24a,Ozawa24,Majumder24}, but as far as we are aware the corresponding model with indirect coupling has not been considered. After introducing the model in Sect. II, our analysis proceeds in three stages. 
\medskip

\noindent 1. Derivation of the continuity equation for the population density in the presence of resetting (Sect. III). As we have recently shown for the Kuramoto model with direct coupling \cite{Bressloff24a}, this is a non-trivial problem since the resulting density equation is itself subject to stochastic resetting. That is, it is not possible to eliminate the stochasticity using a mean field argument (in contrast to local resetting).
\medskip

\noindent 2. Reduction of the infinite-dimensional system to the four dimensional manifold with coordinates $(r(t),R(t),\phi(t),\Phi(t))$ using an OA ansatz (Sect. IV). The validity of this dimensional reduction was previously established in the absence of resetting \cite{Schwab12a}, and thus still holds when resetting is included, provided that the set of initial phases also lies on the OA manifold. A significant feature of the OA dynamics is that only $r(t)$ and $\phi(t)$ reset, which is an example of {\em subsystem resetting}.
We numerically solve the stochastic OA dynamics and explore how the behavior depends on the cell density parameter $\sigma_0$. We also show how subsystem resetting can induce a switch in behavior when the deterministic system operates in a bistable regime. 
\medskip

\noindent 3. Slow/fast analysis of the OA dynamics with subsystem resetting (Sect V). In the particular case $\omega_0=\Omega$, the OA dynamics reduces to a planar system with coordinates $(r(t),R(t))$. In the high density limit $\sigma_0\rightarrow \infty$, we have $R(t)\rightarrow r(t)$ for all $t$ so that $R(t)$ effectively resets when $r(t)$ does. Hence, we recover the one-dimensional OA dynamics of the classical Kuramoto model with global resetting\cite{Sarkar22,Bressloff24a}. On the other hand, in the low density regime ($\sigma_0\ll 1$), $R(t)$ varies much more slowly than $r(t)$ so we can use a separation of time-scales. That is, we first treat $R(t)$ as a constant and calculate the non-equilibrium stationary state (NESS) of the probability density for $r$. The slow variable $R(t)$ is then determined by averaging $r$ with respect to the NESS. If $R(0)=0$ and the resetting rate is sufficiently small, then partial synchrony emerges as $R(t)$ slowly increases with $t$.
\medskip

Before proceeding further, it is helpful to distinguish between our analysis of subsystem resetting and a recent study of the classical Kuramoto model with subsystem resetting\cite{Majumder24}. The latter authors consider a reset protocol in which a chosen proper subset of phase oscillators are simultaneously
reset to a synchronized state at random times, while the remaining oscillators continue from their current state (nonresetting population). It is found that for sufficiently fast resetting, even if only a small fraction of the oscillators reset, the nonresetting population also synchronizes (although not necessarily to a steady state.). This is established using a combination of numerical simulations and a reduction of the nonresetting dynamics to an OA manifold in the continuum limit. In our model all of the oscillators reset under a global resetting protocol, but the external medium does not reset. This then results in an OA dynamics with subsystem resetting.

\setcounter{equation}{0}

\section{Kuramoto model with coupling through an external medium}

\begin{figure}[b!]
\centering
\includegraphics[width=10cm]{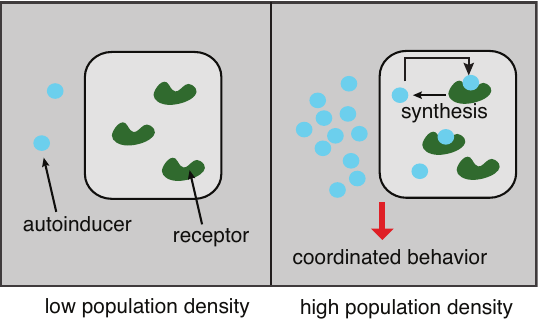} 
\caption{A schematic illustration of bacterial quorum sensing (QS) at the single-cell level. QS involves the production and extracellular secretion of certain signaling molecules known as autoinducers. Each cell also has 
receptors that can detect the autoinducers via ligand-receptor binding. If the probability of detection is sufficiently high then this can activate gene transcription, including the synthesis of the autoinducers. However, since there is a low likelihood of an individual bacterium detecting its own secreted autoinducers, the cell must encounter 
signaling molecules secreted by other cells in its environment in order for gene transcription to be activated. When only a few other bacteria of the same kind are 
in the vicinity (low bacterial population density), diffusion reduces the concentration of the inducer in the surrounding medium to almost zero, resulting in small 
amounts of inducer being available to bind to receptors. On the other hand, as the population grows, the concentration of the inducer passes a threshold, causing more inducer to be 
synthesized. This generates a positive feedback loop such that all of the cells 
initiate transcription at approximately the same time, resulting in some form of collective behavior.}
\label{fig2}
\end{figure}

Consider a population of $\calN$ limit cycle oscillators that are diffusively coupled to an external medium. One important example occurs in a synthetic version of bacterial quorum sensing, in which cells release a signaling molecule known as an autoinducer that diffusively couples a population of genetic oscillators, with the goal of enhancing the oscillatory response via synchronization, see Fig. \ref{fig2}. When the diffusing molecules leave a cell and enter the extracellular environment their concentration is diluted by a factor $\alpha=V_{\rm int}/V_{\rm ext}$, which is the ratio of the cell volume to the total extracellular volume.
Let $z_j(t)$, $j=1,\ldots,\calN$, denote the complex amplitude of the $j$th oscillator at time $t$. The state of the environment is then represented by a complex variable $Z(t)$, which diffusively couples to each $z_j$ with a coupling strength $K$. We also assume that it has an additional oscillatory component of frequency $\Omega$. Let $\omega_j$ be the natural frequency of the $j$-th oscillator. The frequencies $\omega_j$ are randomly drawn from a distribution $h(\omega)$, which is taken to be an even function about a mean frequency $\omega_0$.

Finally,  introducing the dimensionless cell density $\sigma_0=\alpha N$ and moving to a rotating frame with frequency $\omega_0$, the dynamics of the system can be represented by the equations \cite{Schwab12,Schwab12a}
\begin{subequations}
\label{QSz}
\begin{align}
\frac{dz_j}{dt}&=(\lambda_0+i\omega_j-|z_j|^2)z_j-K(z_j-Z),\\
\frac{dZ}{dt} &=\frac{\sigma_0 K}{\calN}\sum_{j=1}^{\calN} (z_j-Z)- i[\omega_0-\Omega]Z.
\end{align}
\end{subequations}
(In this paper we assume that degradation of autoinducer within the external medium is negligible.) 
In the rotating frame the frequencies $\omega_j$ are drawn from an even distribution $g(\omega)=h(\omega-\omega_0)$ with zero mean. 
Note that the density factor $\sigma_0$ in equation (\ref{QSz}b) is the origin of the density-dependent collective response of bacterial QS systems. Intuitively speaking, increasing the effective coupling $ K$ should favor synchronization, whereas having a wider distribution of frequencies $\omega_j$ should favor desynchronization. These opposing effects are well known within the theory of oscillator synchronization \cite{Kuramoto84,Pikovsky03}.

Suppose that Eq. (\ref{QSz}a) is rewritten in polar coordinates with $z_j=r_j\e^{i\theta_j}$ and $Z=R\e^{i\Phi}$:
\begin{subequations}
\begin{align}
\frac{dr_j}{dt}&=(\lambda_0-K-r_j^2)r_j +KR \cos (\Phi-\theta_j),\\
\frac{d\theta_j}{dt}&=\omega_j +\frac{KR}{r_j}\sin(\Phi-\theta_j).
\end{align}
\end{subequations}
The main simplification in the large $\lambda_0$ regime is to note that $r_j\approx \sqrt{\lambda_0}$ in steady-state for all $j=1,\ldots,N$. Thus, we can effectively eliminate the dynamics of the amplitudes $r_j$. After performing the rescalings $r_j \rightarrow r_j/\sqrt{\lambda_0}$ and $R\rightarrow R/\sqrt{\lambda_0}$, equations (\ref{QSz}) reduce to \cite{Schwab12a}
\begin{subequations}
\label{Kur}
\begin{align}
\frac{d\theta_j}{dt}&=\omega_j+KR\sin (\Phi-\theta_j),\\
\frac{dZ}{dt}&= \sigma_0 K({z}(t)-Z(t))- +i[\omega_0-\Omega]Z(t),
\end{align}
\end{subequations}
with
\begin{equation}
z(t):=\frac{1}{\calN}\sum_{j=1}^{\calN}\e^{i\theta_j(t)}=r(t)\e^{\phi(t)}.
\end{equation}
This is a modified version of the classical Kuramoto model where $Z$ is identified with the state of an external medium rather than the first circular moment $ z(t)$ of the $N$ oscillators. The latter has a geometric interpretation as the centroid of the phases with $ \phi(t)$ equal to the average phase and $r(t)$ a measure of the degree of
phase-coherence.
A completely incoherent state corresponds to the case $r=0$, whereas a completely synchronized state satisfies $r=1$.

In this paper we consider a stochastic version of the Kuramoto model (\ref{Kur}) in which each oscillator simultaneously resets to its initial phase $\theta_{0,j}$ at a fixed Poisson rate $\lambda$. On the other hand, the external complex amplitude $Z(t)$ does not reset. Following Refs. \cite{Nagar23,Bressloff24a}, we distinguish this global update rule from a local resetting scheme in which the oscillators independently reset. Let $T_{n}$ denote the $n$th resetting time with $n\geq 1$ and $T_0=0$. The inter reset times $ {\tau}_n ={\calT}_n- {\calT}_{n-1}$ are exponentially distributed with
\begin{equation}
 \P[ {\tau}_n\in [s,s+ds]]=\lambda \e^{-\lambda s}ds
\end{equation} 
 In addition, the number of resets occurring in the time interval $[0,t]$ is given  by the Poisson process 
 \begin{align}
 {N}(t)=n,\quad  {\calT}_{n}\leq t < {\calT}_{n+1},\quad \P[ {N}(t)=n]=  \frac{(\lambda t)^{n}\e^{-\lambda t}}{n!},
\label{Pois2}
\end{align}
with $\E[N(t)]=\lambda t=\mbox{Var}[N(t)]$. Note that $N(t)$ is defined to be right-continuous: $N (\calT_n^-)=n-1$ whereas $N(\calT_n)=n$

As we have discussed elsewhere \cite{Bressloff24a}, instantaneous global resetting can be implemented using the stochastic calculus of jump processes. In particular, introducing the differential
\begin{equation}
dN(t)=h(t)dt,\quad h(t)=\sum_{n\geq 1}\delta(t-\calT_n),
\end{equation}
Eqs. (\ref{Kur}) become
\begin{subequations}
\label{Kurres}
\begin{align}
 \frac{d\theta_j(t)}{dt}&=\omega_j+K R(t) \sin(\Phi(t)-\theta_j(t))   +h(t)(\theta_{j,0}-\theta_j(t^-)),\\
\frac{dZ}{dt}&= \sigma_0 K({z}(t)-Z(t))- i[\omega_0-\Omega]Z(t),
\end{align}
\end{subequations}
Integrating Eq. (\ref{Kurres}a) with respects to $t$ implies that
\begin{align}
 \theta_j(t) &=\theta_{j,0}+\omega_jt+K \int_0^t  R(s) \sin(\Phi(s)-\theta_j(s)) ds 
 +\sum_{m=1}^{N(t)}(\theta_{j,0}-\theta_j(T_{m}^-)).
 \end{align}
Setting $t= {\calT}_n$ and $t= {\calT}_n^-$, respectively, and subtracting the resulting pair of equations shows that $\theta_j( {\calT}_n)= \theta_j( {\calT}_{n}^-)=\theta_{j,0} - \theta_j( {\calT}_{n}^-)=\theta_{j,0} $
Hence, we recover instantaneous resetting. A crucial element in establishing this equivalence is the right-continuity of the Poisson process $N(t)$.

An alternative resetting protocol would be for the external medium to reset rather than the phase oscillators. (One could also consider a combination of both resetting protocols.) That is, $Z(\calT_n^-)\rightarrow Z_0$. Eq. (\ref{Kurres}b) then becomes
\begin{align}
\label{Kurres1}
\frac{dZ}{dt}&=\sigma_0 K({z}(t)-Z(t))- i[\omega_0-\Omega]Z(t)  +h(t)(Z_0-Z(t^-)).
\end{align}
In terms of applications to quorum sensing, taking $Z_0=0$ would represent the clearing of all autoinducers from the external environment. This assumes that all oscillators are intrinsic in the sense that they do not require external coupling in order to oscillate. In this paper we assume that $Z(t)$ does not reset. (This ultimately leads to dynamics on the OA manifold that is subject to subsystem resetting, see Sect. IV.)

\setcounter{equation}{0}

\section{Derivation of global density equation}

A well-known method for analyzing the classical Kuramoto model is to consider the PDE for the joint phase and natural frequency density in the continuum limit $\calN\rightarrow \infty$ \cite{Sakaguchi88,Strogatz91,Crawford94,Crawford99,Strogatz00}. 
In most cases the density equation is simply written down in the form of a Liouville equation, or a Fokker-Planck equation in the presence of phase fluctuations. However, there is an implicit assumption that the continuum or mean field limit exists. 
The latter has been proven rigorously by treating the noisy Kuramoto model as a system of Brownian particles on an $\calN$-torus\cite{Dai96}, and this result has recently been extended to a wider class of interacting particle systems on the torus \cite{Carrillo20,Pavliotis21}. In a previous paper we exploited the connection between the classical Kuramoto model and systems of interacting Brownian particles on a torus to incorporate the effects of stochastic resetting into the corresponding continuum model \cite{Bressloff24a}. In this section we use an analogous construction to incorporate global resetting into the density equation for the Kuramoto model coupled via an external medium.

Introduce the global density or empirical measure
\begin{align}
\rho^{(\calN)}(\theta,t,\omega)&\equiv \frac{1}{\calN}\sum_{j=1}^{\calN}\rho_j(\theta,t,\omega) 
=\frac{1}{\calN}\sum_{j=1}^{\calN}\delta (\theta-\theta_j(t))\delta(\omega-\omega_j).
\end{align}
Consider an arbitrary smooth test function $f: [0,2\pi]\times  \R\rightarrow \R$.
Setting 
\begin{align}
&\frac{1}{\calN}\sum_{j=1}^{\calN}f(\theta_j(t),\omega_j)=  \int_0^{2\pi}d\theta \,\int_{\R}d\omega \,  \rho^{(\calN)}(\theta,t,\omega)f(\theta,\omega),
\end{align}
 we have
\begin{align}
 & \int_0^{2\pi}d\theta \,\int_{\R}d\omega \,  f(\theta,\omega)\frac{\partial \rho^{(\calN)}(\theta,t,\omega)}{\partial t}   =\frac{1}{\calN}\sum_{j=1}^{\calN} \frac{df(\theta_j(t),\omega_j)}{dt},
 \end{align}
 with
 \begin{align}
\frac{df}{dt}(\theta_j(t),\omega_j)  =   \frac{\partial f}{\partial \theta_j}(\theta_j(t), \omega_j) \frac{d \theta_j(t)}{dt} 
  &=  \bigg [ \omega_j+ KR(t) \sin(\Phi(t)-\theta_j(t))\bigg ]\frac{\partial f}{\partial \theta_j}(\theta_j(t), \omega_j)\nonumber \\
 &\quad + h(t ) \bigg [f(\theta_{0,j},\omega_j)-f(\theta_j(t^-),\omega_j)\bigg ]    
\end{align}
Using the definition of $\rho^{(\calN)}$ we have 
\begin{align}
 \int_0^{2\pi}d\theta \, \int_{\R}d\omega f(\theta,\omega)\frac{\partial \rho^{(\calN)}(\theta,t,\omega)}{\partial t}  
&=\int_0^{2\pi}d\theta \,\int_{\R}d\omega\, \rho^{(\calN)}(\theta,t,\omega)\bigg [  \calV[\theta,\omega,Z]\partial_{\theta} f(\theta,\omega)\bigg ]\\
 &\quad +h(t)\int_0^{2\pi}d\theta \,\int_{\R}d\omega\, [\rho^{(\calN)}_0(\theta,\omega)-\rho(\theta,\omega,t^-)]f(\theta,\omega),\nonumber
\end{align}
where
\begin{equation}
\calV[\theta,\omega,Z]=\omega+\frac{K}{2 i}(Z(t)\e^{-i\theta}-Z^*(t) \e^{i\theta}),
\label{calV}
\end{equation}
and 
\begin{equation}
\label{shot}
\rho_0^{(\calN)}(\theta,\omega)= \frac{1}{\calN}\sum_{j=1}^{\calN} \delta(\theta-\theta_{0,j})\delta(\omega-\omega_j).
\end{equation}
Integrating by parts the various terms involving derivatives of $f$ and using the fact that $f$ is arbitrary yields the following equation for $\rho$ (in the weak sense):
\begin{align}
  \label{SPDE}
 \frac{\partial \rho^{(\calN)}(\theta,t,\omega)}{\partial t}&= -\frac{\partial }{\partial \theta}\bigg [\rho^{(\calN)}(\theta,t,\omega)  \calV[\theta,\omega,Z]
  \bigg ]  +h(t) [\rho^{(\calN)}_0(\theta,\omega)-\rho^{(\calN)}(\theta,t^-,\omega)],
\end{align}
where $Z(t)$ evolves according to equation (\ref{Kurres}b) with
\begin{equation}
\label{zbar0}
{z}(t)=\langle \e^{i\theta} \rangle_{\rho} \equiv \int_{-\infty}^{\infty} \int_0^{2\pi}\rho^{(\calN)}(\theta,\omega,t)\e^{i\theta}d\theta d\omega.
\end{equation}

If $\calN$ is finite, then Eq. (\ref{SPDE}) is highly singular since the solution is a sum of Dirac delta functions. There are two ways to obtain a non-singular PDE. The first is to take expectations with respect to the Poisson process $N(t)$ that generates the sequence of resetting times. However, as we show below, this leads to a moment closure problem even in the continuum limit. The second is to take the continuum limit but keep stochastic resetting, which is the approach we take in Sect. IV. In order to understand how a smooth density can arise in the continuum limit, consider an arbitrary smooth function $f(\omega)$ and the identities
\begin{align*}
\frac{1}{\calN}\sum_{j=1}^{\calN}f(\omega_j)
 =\int_{\R}\rho^{(\calN)}(\omega)f(\omega)d\omega
\end{align*}
where
$\rho^{(\calN)}(\omega)={\calN}^{-1}\sum_{j=1}^{\calN}\delta(\omega-\omega_j)$, 
and
\begin{align*}
 \lim_{\calN \rightarrow \infty}\frac{1}{\calN}\sum_{j=1}^{\calN}f(\omega_j)&
 =\int_{\R}g(\omega)f(\omega)d\omega
\end{align*}
It follows from the arbitrariness of $f$ that
\[ \lim_{\calN \rightarrow \infty}\rho^{(\calN)}(\omega)=g(\omega).\]
in the sense of distributions.

In order to take expectations of Eq (\ref{SPDE}) with respect to $N(t)$ we use the following results. Since $\rho^{(\calN)}(\theta,T_n^-,\omega)$ only depends on  jump times prior to the $n$th resetting, we have
\begin{align}
\E[h(t) [\rho_0^{(\calN)}(\theta,\omega)-\rho^{(\calN)}(\theta,t^-,\omega)]] 
&=\E[\rho_0^{(\calN)}(\theta,\omega)-\rho^{(\calN)}(\theta,t^-,\omega)]\E[h(t)]\nonumber\\
&=[\rho_0^{(\calN)}(\theta,\omega)-p(\theta,t,\omega)]\E[h(t)],
\end{align}
where $p(\theta,t,\omega)=\E[\rho^{(\calN)}(\theta,t,\omega)]$. Moreover,
$N(t)-N(\tau)=\int_{\tau}^tdN(s)$ so that $\int_{\tau}^t\E[dN(s)] = \E[N(t)-N(\tau)]=\lambda(t-\tau)$ and so
$\E[dN(t)]=\lambda dt$. That is, $\E[h(t)]=\lambda$. Hence, substituting for $\calV$ using Eq. (\ref{calV}) and then taking expectations of Eq. (\ref{SPDE}) gives
\begin{subequations}
  \label{SPDEav}
\begin{align}
 \frac{\partial p(\theta,t,\omega)}{\partial t}&=\lambda[\rho_0^{(\calN)}(\theta,\omega)-p(\theta,t,\omega)] 
  -\frac{\partial }{\partial \theta}\bigg \{\omega p(\theta,t,\omega) +\frac{K}{2 i}\e^{-i\theta}\E[\rho^{(\calN)}(\theta,t,\omega)Z(t)]\nonumber \\
&\qquad -\frac{K}{2 i}\e^{i\theta}\E[\rho^{(\calN)}(\theta,t,\omega)\overline{Z}(t)]  \bigg \} .\nonumber
\end{align}
On the other hand, taking expectations of equations (\ref{Kurres}b) and (\ref{zbar0}), we have
\begin{align}
\frac{d\calZ}{dt}&= \sigma_0 K(\E[z(t)]-\calZ(t))- i[\omega_0-\Omega]\calZ(t),
\end{align}
where $\calZ(t)=\E[Z(t)]$ and
\begin{equation}
\label{zbar0p}
\E[z(t)]=\langle \e^{i\theta} \rangle_{p} \equiv \int_{-\infty}^{\infty} \int_0^{2\pi}p(\theta,\omega,t)\e^{i\theta}d\theta d\omega.
\end{equation}
\end{subequations}

Eq. (\ref{SPDEav}a) does not form a closed system since the one-point density $p(x,t,\omega)=\E[\rho^{(\calN)}(\theta,t,\omega)]$ couples to the two-point density $\E[\rho^{(\calN)}(\theta,t,\omega)Z(t)]$ and its complex conjugate. We are using the fact that $Z(t)$ is a linear functional of $\rho(\theta,t,\omega)$, which follows from Eqs. (\ref{Kurres}b) and (\ref{zbar0}). Unfortunately, we cannot appeal to a mean field limit in which $\E[\rho^{(\calN)}(\theta,t,\omega)Z(t)]\rightarrow p(\theta,t,\omega)\calZ(t)$, which means that we cannot eliminate the stochasticity by averaging. Hence, taking the continuum limit of Eq. (\ref{SPDE}) we have
\begin{align}
  \label{SPDE2}
 \frac{\partial \rho(\theta,t,\omega)}{\partial t}&= -\frac{\partial }{\partial \theta}\bigg [\rho(\theta,t,\omega)  \calV[\theta,\omega,Z]
  \bigg ] +h(t) [\rho_0(\theta,\omega)-\rho(\theta,t^-,\omega)],
\end{align}
with
\begin{align}
\label{rho0}
\rho_0(\theta,\omega)&=  \lim_{\calN\rightarrow \infty}\frac{1}{\calN}\sum_{j=1}^{\calN} \delta(\theta-\theta_{0,j})\delta(\omega-\omega_j).
\end{align}

\setcounter{equation}{0}
\section{Stochastic dynamics on the Ott-Antonsen manifold}

The next step in our analysis ts to project Eq. (\ref{SPDE2}) onto the OA manifold along analogous lines to the deterministic model of Ref. \cite{Schwab12a}. First, note that the continuum density satisfies the normalization condition
\begin{equation}
\int_{0}^{2\pi}\rho(\theta,\omega,t)d\theta =\lim_{\calN\rightarrow \infty}\frac{1}
{\calN}\sum_{j=1}^{\calN}\delta(\omega-\omega_j)= g(\omega).
\end{equation}
It follows that $\rho$ can be expanded as a Fourier series in $\theta$:
\begin{align}
\rho(\theta,\omega,t)=\frac{g(\omega)}{2\pi}\left (1+\sum_{n=1}^{\infty}\left [\rho_n(\omega,t)\e^{in\theta}+\mbox{c. c.}\right ]\right ).
\end{align}
An analogous Fourier expansion holds for the reset density:
\begin{align}
\rho_0(\theta,\omega)
&=  \frac{g(\omega)}{2\pi}\left (1+\sum_{n=1}^{\infty}\left [\rho_{0,n}(\omega)\e^{in\theta}+\mbox{c. c.}\right ]\right ).
\end{align}
As it stands, solving the initial value problem for $\rho$ involves solving an infinite hierarchy of equations for the coefficients $\rho_n$. However, a major simplification can be achieved by applying the OA ansatz $\rho_n =\eta^n$ so that
\begin{align}
\label{OA}
\rho(\theta,\omega,t)=\frac{g(\omega)}{2\pi}\left (1+\sum_{n=1}^{\infty}\left [\eta^n(\omega,t)\e^{in\theta}+\mbox{c. c.}\right ]\right ).
\end{align}
Numerical simulations of the full system in the absence of resetting indicate that the dynamics is captured by such a reduction \cite{Schwab12a}. We add the further constraint that the reset density also lies on the OA manifold:
\begin{equation}
\label{OA0}
\rho_0(\theta,\omega)=\frac{g(\omega)}{2\pi}\left (1+\sum_{n=1}^{\infty}\left [r_0^n\e^{in\theta}+\mbox{c. c.}\right ]\right ).\end{equation}
For simplicity, the OA coefficient $r_0$ is taken to be real and independent of $\omega$. It then follows that
$\rho_0(\theta,\omega)=g(\omega)p_0(\theta)$ where $p_0(\theta)$ is the $\omega$-independent distribution of initial phases:
\begin{align}
p_0(\theta)&\equiv  \lim_{\calN\rightarrow \infty}\frac{1}{\calN}\sum_{j=1}^{\calN} \delta(\theta-\theta_{0,j}) 
=\frac{1}{2\pi}\left (-1+\sum_{n=0}^{\infty}\left [r_0^n\e^{in\theta}+\mbox{c. c.}\right ]\right )\nonumber \\
&=\frac{1}{2\pi}\left (-1+\frac{1}{1-r_0\e^{i\theta}}+\frac{1}{1-r_0\e^{-i\theta}}\right ) 
=\frac{1}{2\pi}\left (-1+\frac{2-2r_0\cos \theta}{1-2r_0\cos \theta+r_0^2} \right )\nonumber \\
&=\frac{1}{2\pi}\left (\frac{1-r_0^2}{1-2r_0\cos \theta+r_0^2} \right ).
\end{align}

Substituting Eqs.(\ref{OA}) and (\ref{OA0}) into the continuity Eq. (\ref{SPDE}) gives
\begin{align}
&\eta^{n-1}(\omega,t)\bigg [\frac{\partial \eta}{\partial t}+i\omega \eta +\frac{K}{2}(Z\eta^2-Z^*)\bigg ]
=h(t) [\eta_0^n(\omega)-\eta^n(\omega,t^-)] ,
\end{align}
with $Z(t)$ evolving according to equation (\ref{Kurres}b).
In other words, between resetting events we have
\begin{equation}
\frac{\partial \eta}{\partial t}+i\omega \eta +\frac{K}{2}(Z\eta^2-Z^*)=0,
\label{al}
\end{equation}
which is supplemented by the reset condition
\begin{equation}
\eta(\omega,\calT_n^-)\rightarrow r_0
\end{equation} 
at the reset times $\calT_n$, $n \geq 1$.
Moreover, Eq. (\ref{zbar0}) implies that on the OA manifold
\begin{equation}
\label{zbar}
{z}(t)=\int_{-\infty}^{\infty}g(\omega)\eta^*(\omega,t)d\omega.
\end{equation}

\begin{figure}[t!]
\centering
\includegraphics[width=16cm]{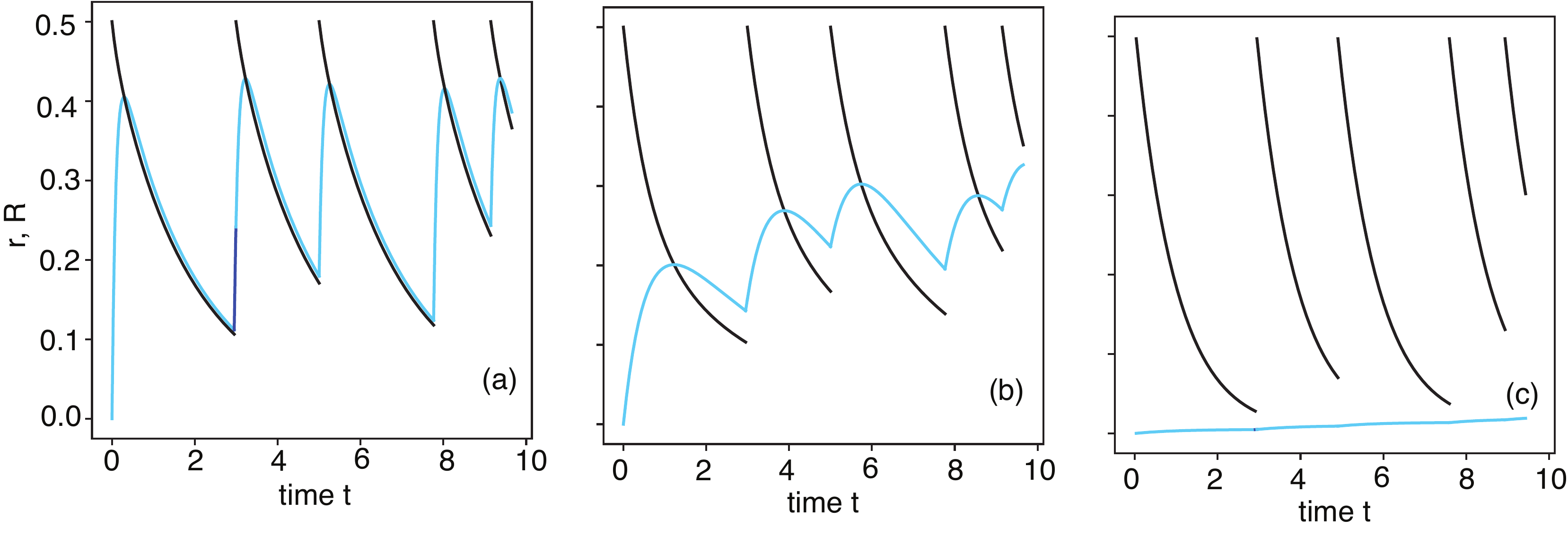} 
\caption{Planar dynamics (\ref{OArR2}) on the OA manifold with resetting for $\omega_0=\Omega$ and different values of the density: (a) $\sigma_0K=10$, (b) $\sigma_0K=1$, (c) $\sigma_0K=0.01$. Other parameters are $r_0=0.5$, $\lambda=0.5$, $\Delta=1$, and $K=1$. Black (dark) curves show $r(t)$ and blue (light) curves show $R(t)$.}
\label{fig3}
\end{figure}

\begin{figure}[t!]
\centering
\includegraphics[width=16cm]{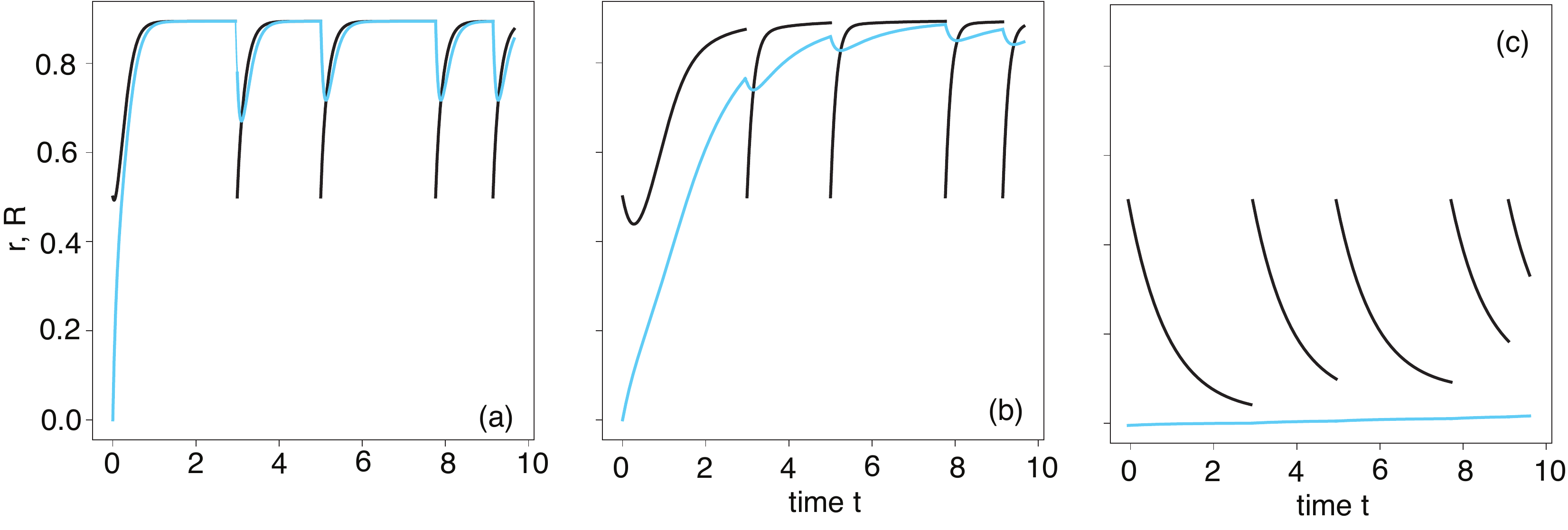} 
\caption{Same as Fig. \ref{fig3} except that $K=10$.}
\label{fig4}
\end{figure}

Further simplification can be achieved in the case of a Lorentzian distribution of frequencies:
\begin{equation}
\label{Lor}
g(\omega)=\frac{1}{\pi}\frac{\Delta}{\omega^2+\Delta^2}.
\end{equation}
Substituting (\ref{Lor}) into Eq. (\ref{zbar}) and evaluating the resulting contour integral establishes that
\[{z}(t)=\eta^*(-i\Delta,t).\]
The resetting rule then becomes
$z(t)\rightarrow r_0 $.
Hence, setting $\omega=-i\Delta$ in Eq. (\ref{al}) and taking the complex conjugate yields the following equations for $z$ and $Z$ between resetting events:
\begin{subequations}
\label{zZ}
\begin{equation}
\frac{d {z}}{d t}+\Delta {z}+\frac{K}{2}(Z^*z^2-Z)=0.
\end{equation}
\begin{equation}
\frac{dZ}{dt}= {\sigma_0K}  \left ( {z} -Z\right )- i[\omega_0-\Omega]Z,
\end{equation}
\end{subequations}
 Introducing the planar coordinates $Z(t)=R(t)\e^{i\Phi(t)}$ and $ z(t)=r(t)\e^{i\phi(t)}$ gives
\begin{subequations}
\begin{equation}
\frac{dr}{dt}+ir\frac{d\phi}{dt}+\Delta r+\frac{KR}{2}\left (r^2\e^{i[\phi-\Phi]} -\e^{-i[\phi-\Phi]}\right )=0.
\end{equation}
and
\begin{equation}
\frac{dR}{dt}+iR\frac{d\Phi}{dt}= {\sigma_0K}  \left (r\e^{i[\phi-\Phi]} -R\right )- i[\omega_0-\Omega]R,
\end{equation}
\end{subequations}
supplemented by the subsystem resetting rule ($R,\Phi$ do not reset)
\begin{equation}
\label{proto}
r(t)\rightarrow r_0>0,\quad \phi(t)\rightarrow 0.
\end{equation}
Equating real and imaginary parts yields
\begin{subequations}
\label{OArR}
\begin{align}
\frac{dr}{dt}&=-\Delta r+\frac{KR\cos[\Phi-\phi]}{2}( 1-r^2),\\
\frac{dR}{dt}&= {\sigma_0K}   r\cos[\Phi-\phi] -\sigma_0K R,
\end{align}
and
\begin{align}
r\frac{d\phi}{dt}&=\frac{KR\sin (\Phi-\phi)}{2}(1+r^2),\\
R\frac{d\Phi}{dt}&=-\sigma_0Kr\sin(\Phi-\phi)-[\omega_0-\Omega] R.
\end{align}
\end{subequations}

\begin{figure*}[t!]
\centering
\includegraphics[width=14cm]{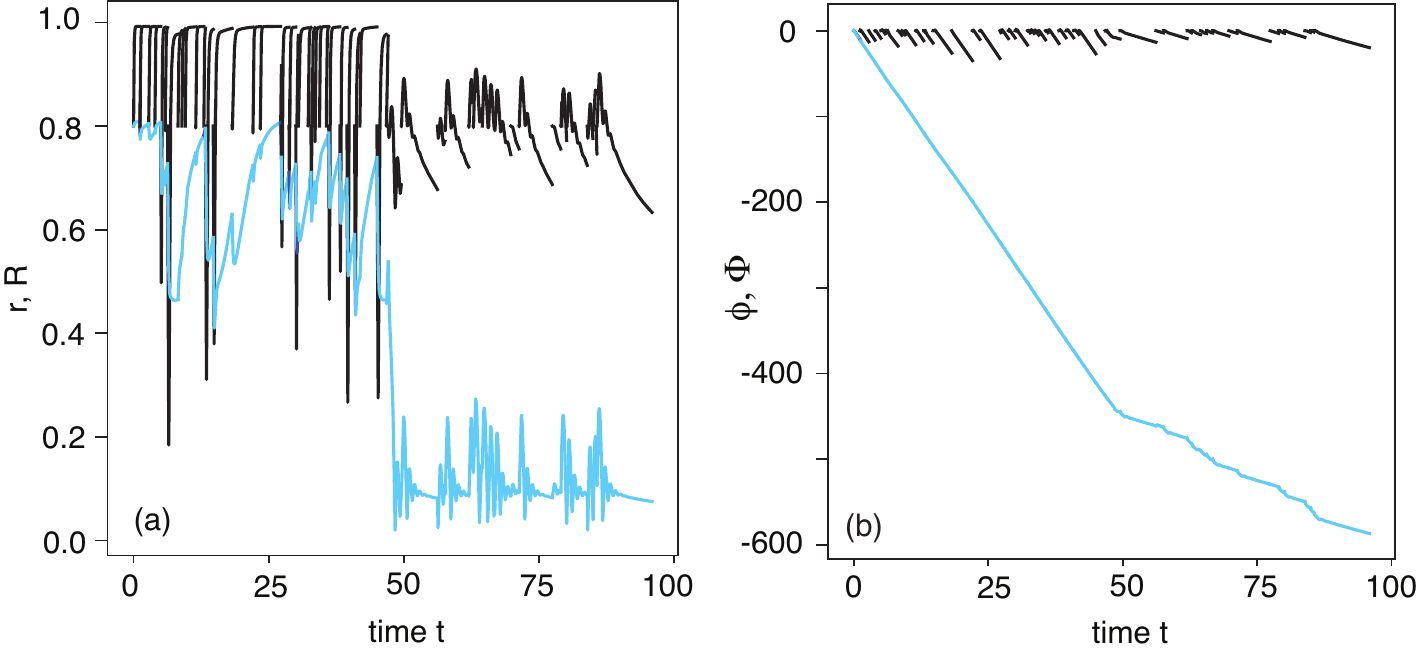} 
\caption{Noise-induced transition between a pair of coherent states for the full OA system given by Eqs. (\ref{proto}) and (\ref{OArR}a-d) with $r_0=0.8$. Other parameter values are $\omega_0-\Omega =10$, $\Delta=0.1$, $K=20$, $\lambda=0.5$, and $\sigma_0=0.05$. (a) Plots of $r(t)$ (dark curves) and $R(t)$ (light curves) as a function of $t$. (b) Corresponding plots of $\phi(t)$ (dark curves) and $\Phi(t)$ (light curves).}
\label{fig5}
\end{figure*}

\begin{figure*}[t!]
\centering
\includegraphics[width=14cm]{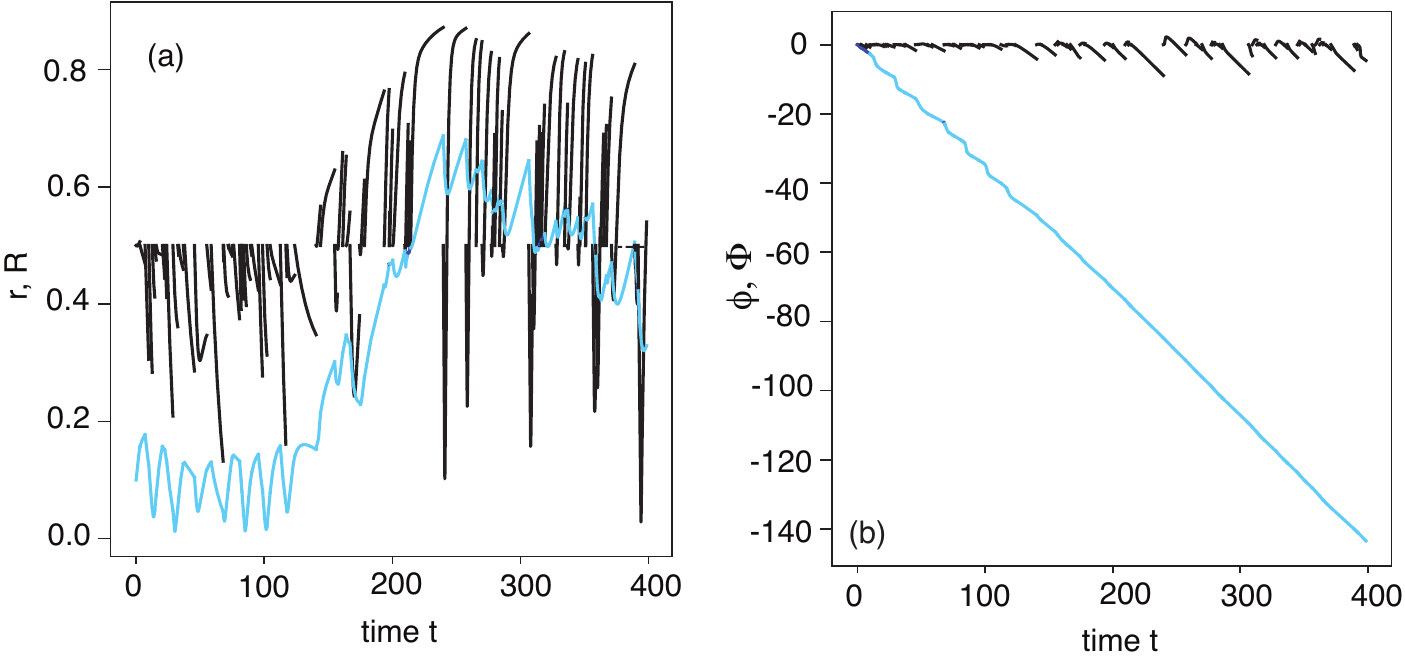} 
\caption{Noise-induced transition from an incoherent state to a coherent states for the full OA system given by Eqs.  Eqs. (\ref{proto}) and (\ref{OArR}a-d) with $r_0=0.5$. Other parameter values are $\omega_0-\Omega =0.4$, $\Delta=0.1$, $K=1.2$,  $\lambda=0.2$, and $\sigma_0=0.05$. (a) Plots of $r(t)$ (dark curves) and $R(t)$ (light curves) as a function of $t$. (b) Corresponding plots of $\phi(t)$ (dark curves) and $\Phi(t)$ (light curves).}
\label{fig6}
\end{figure*}

We identify the variables $(r(t),R(t),\phi(t),\Phi(t))$ as the coordinates of a four-dimensional OA manifold. In the absence of resetting ($\lambda= 0$), this can be reduced to an effective three-dimensional manifold. That is,
dividing Eqs. (\ref{OArR}c,d) by $r$ and $R$ respectively, and subtracting the results yields the following equation for the phase difference $\psi=\Phi-\phi$:
\begin{equation}
\label{psi}
\frac{d\psi}{dt}=\Omega -\omega_0-K\sin \psi \left (\frac{r\sigma_0}{R} +\frac{R}{2r}[1+r^2] \right ).
\end{equation}
Eqs. (\ref{OArR}a,b) and (\ref{psi}) with $\lambda=0$ form a system of three coupled equations for the variables $r(t),R(t),\psi(t)$. 
The resulting deterministic dynamics and associated phase diagrams were analyzed extensively in Ref. \cite{Schwab12a}. However, extending the analysis to include the effects of resetting is non-trivial. First, the resetting protocol (\ref{proto}) requires dealing with the full 4-dimensional state space. Second, it is difficult to solve the corresponding PDE for the probability density $p(r,R,\phi,\Phi,t)$.

Further simplification can be achieved under the additional condition $\Phi=\omega_0$. In that case $\psi=0$ is a stable mode so that if $\phi(0)=\Phi(0)=0$ then they remain zero for all $t$. Setting $\psi=0$ in Eqs. (\ref{OArR}a,b) then leads to the planar dynamical system
\begin{subequations}
\label{OArR2}
\begin{align}
\frac{dr}{dt}&=-\Delta r+\frac{KR}{2}( 1-r^2)+h(t)[r_0-r(t^-)],\\
\frac{dR}{dt}&= {\sigma_0K}  [ r - R].
\end{align}
\end{subequations}
In the absence of resetting, the fixed points $(\overline{r},\overline{R})$ are determined by the conditions $\,\overline{R}=\overline{r}$ and 
$\overline{r}(K-2\Delta -K\overline{r}^2)=0$. The latter is identical to the fixed point condition obtained for the classical Kuramoto model. That is, a stable incoherent state $\overline{r}=0$ loses stability at the critical coupling $K_c=2\Delta $ and a stable partially coherent state coexsist when $K>K_c$.

Example plots of solutions to Eqs. (\ref{OArR2}a,b) with resetting are shown in Figs. \ref{fig3} and \ref{fig4}. We use the same random reset times in each plot but vary $\sigma_0$ and $K$ while the other parameters are fixed. In particular $r_0=0.5$ and $R(0)=0$. Fig. \ref{fig3} shows the case $K<K_c$ for which the steady-state $\overline{r}=0<r_0$ in the absence of resetting. In the high density regime ($\sigma_0K\gg \Delta$) we see that $R(t)\approx r(t)$ between resetting events and $R(t) \rightarrow r_0$ almost instantaneously following each reset. Reducing $\sigma_0K$ to values of $O(\Delta)$ significantly alters the dynamics of $R(t)$ but only weakly modifies $r(t)$. On the other hand, in the low density regime $\sigma_0K \ll \Delta$, there is a significant change in the dynamics of $r(t)$ due to the fact that $R(t)$ is slowly varying with $R(0)=0$. This greatly reduces the initial level of synchronization. The latter is particularly significant when $K>K_c$ and $\overline{r} >r_0>0$, see Fig. \ref{fig4}. However, synchrony is recovered on longer time scales as  we show in Sect. V.

If $\omega_0\neq \Omega$, then it is necessary to solve the full system of Eqs. (\ref{OArR}). In the absence of resetting, one
finds numerically that the low density regime exhibits bistability between either (i) an incoherent state and a coherent state or (ii) two coherent states with different amplitudes $R$ \cite{Schwab12a}. Such bistability persists in the presence of resetting. However, in certain cases, the stochasticity due to subsystem resetting can induce a state transition, which is illustrated in Figs. \ref{fig5} and \ref{fig6}. In both cases we choose parameter values that lie in one of the bistability regions displayed in Fig. 1 of Ref. \cite{Schwab12a}. Fig. \ref{fig5} shows a noise-induced transition between two coherent states $(r,R)\approx (1,0.8)$ and $(r,R)\approx (0.5,0.1)$. Similarly, Fig. \ref{fig6} shows a noise-induced transition from an incoherent state to a coherent state: $(r,R)\approx (0.0,0.0)$ and $(r,R)\approx (0.9.0.82)$. Meanwhile, the environmental phase varies as $\Phi(t)\approx (\omega-\Omega)t$ whereas $\phi(t)$ resets to zero.

The noise-induced transitions shown in Figs. \ref{fig5} and \ref{fig6} are significant in the sense that the underlying mechanism is based on subsystem resetting alone. The standard paradigm is to consider how stochastic resetting affects noise-induced escape or first passage time events that are driven by an additional external noise source such as Brownian motion \cite{Evans20}. Here there are no additional sources of noise. The crucial observation is that only a subset of the variables, namely $(r(t),\phi(t))$, are reset. If all of the variables reset to their initial conditions, then there would be no mechanism for the system to jump from one basin of attraction to another. On the other hand, under subsystem resetting we can interpret the state $(r_0,R(\calT_n),\phi_0,\Phi(\calT_n))$ immediately following the $n$th resetting event as a new initial condition. Finally, it should be noted that the state transitions in Figs. \ref{fig5} and \ref{fig6} appear to be irreversible.

\setcounter{equation}{0}
\section{Slow/fast analysis of the OA dynamics}

Further analysis can be performed on the planar system (\ref{OArR2}) in the low and high density regimes using a slow/fast analysis.

\subsection{Low density limit}

In the low density limit $\sigma_0\rightarrow 0^+$ the external amplitude $R(t)$ is a slow variable whereas the order parameter $r(t)$ is a fast variable. As a first approximation we can treat the factor $R$ appearing in Eq. (\ref{OArR2}a) as a constant $R_0$ so that we obtain 
\begin{align}
\label{SHS}
\frac{dr}{dt}&= -f(r)+h(t)[r_0-r(t^-)], 
\end{align}
with
\begin{align}
f(r)&= \Delta r-\frac{KR_0}{2}( 1-r^2)\nonumber \\
&=\frac{KR_0}{2}\left (r^2+2\Lambda r-1 \right ),\quad \Lambda = \frac{\Delta}{KR_0}.
\label{fL}
\end{align}
The probability density $q(r,t)$ corresponding to a piecewise deterministic equation of the form (\ref{SHS}) satisfies the PDE
\begin{equation}
\frac{\partial q(r,t)}{\partial t}=\frac{\partial [f(r)q(r,t)]}{\partial r}-\lambda q(r,t)+\lambda \delta(r-r_0).
\end{equation}
There exists a unique nonequilibrium stationary state (NESS) $q^*(r)=\lim_{t\rightarrow\infty}q(r,t)$ with
\begin{equation}
f(r)\frac{dq^*(r)}{d r}+(f'(r)-\lambda) q^*(r)=-\lambda \delta(r-r_0).
\label{qr}
\end{equation}
Solving Eq. (\ref{qr}) on either side of $r_0$ gives
\begin{align}
q^*(r)=\left \{ \begin{array}{cc} \frac{\displaystyle A_-}{\displaystyle F(r)} &  \quad r<r_0 \\ & \\ \frac{\displaystyle A_+}{\displaystyle F(r)} & r>r_0\end{array}
\right . ,
\end{align}
where $F(r)$ is the integrating factor
\begin{align}
F(r):&=\exp\left (\int \frac{f'(s)-\lambda}{f(s)}ds\right )= f(r) \exp\left (-\int \frac{\lambda}{f(s)}ds\right ).
\label{Fir}
\end{align}
Second, integrating Eq. (\ref{qr}) over the interval $ [r_0-\epsilon, r_0+\epsilon]$ and taking the limit $\epsilon \rightarrow 0^+$
implies that
\begin{equation}
A_+ -A_-=-\frac{\lambda F(r_0)}{f(r_0)}.
\end{equation}
This equation together with the normalization condition
\begin{equation}
\int_0^1q^*(r)dr=1
\end{equation}
determine the unknown coefficients $A_{\pm}$. 

\begin{figure*}[t!]
\centering
\includegraphics[width=14cm]{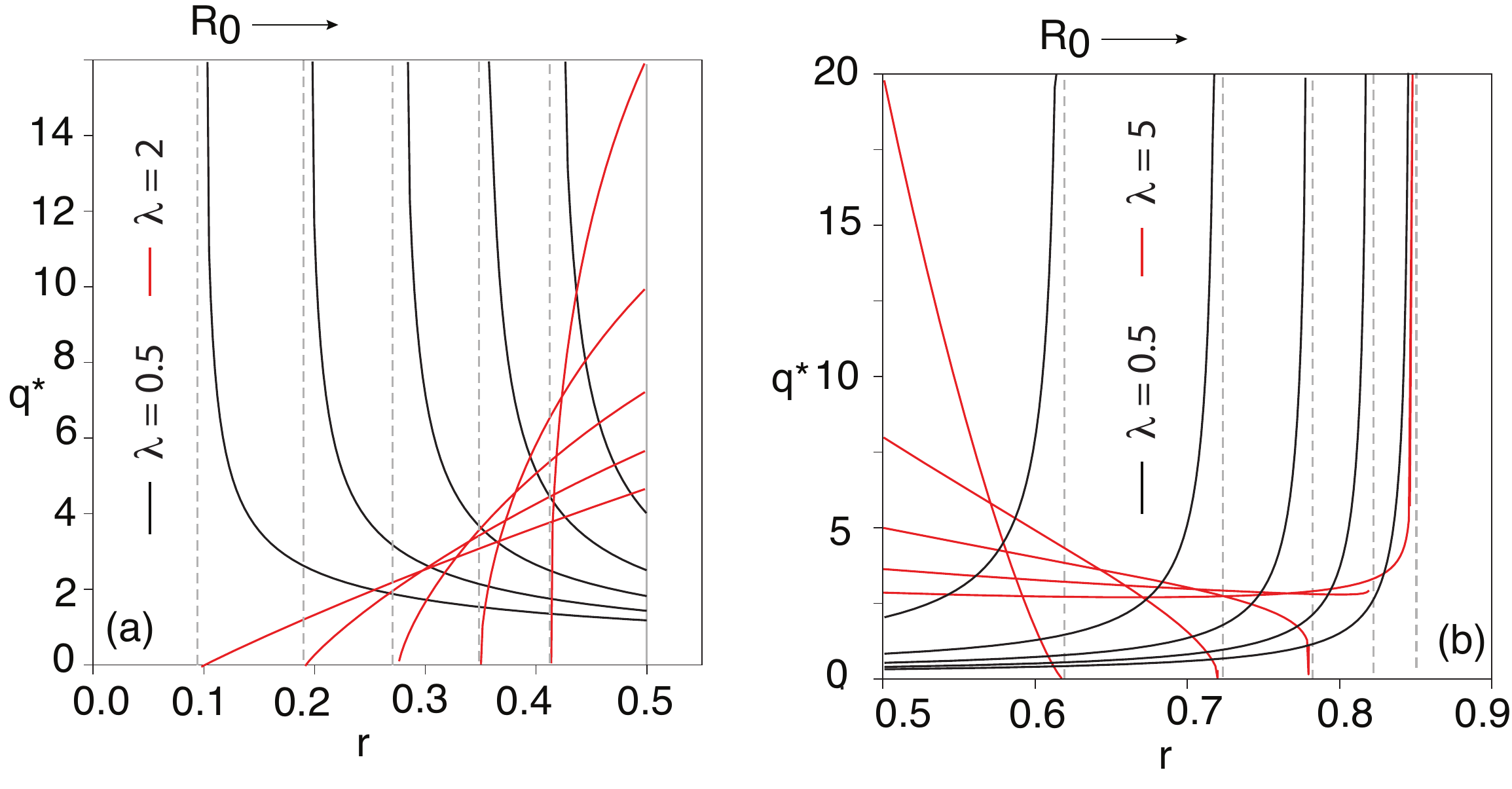} 
\caption{Plots of the NESS $q^*(r)$ given by Eq. (\ref{q1}) as a function of the order parameter $r$ for different values of $R_0$. (a) $r_+ <r_0 =0.5$ for $R_0=0.2, 0.4,0.6,0.8,1.0$ and $\lambda=0.5$ (black curves) or $\lambda=2$ (red curves). (b) $0.5 =r_0 <r_+ $ for $R_0=2,3,4,5,6$ and $\lambda=0.5$ (black curves) or $\lambda=5$ (red curves). Other parameters are $\Delta =1$ and $K=1$.} 
\label{fig7}
\end{figure*}

First consider the case of no resetting. Factorizing $f(r)$ we have the deterministic equation
\begin{equation}
\frac{dr}{dt}=-\frac{KR_0}{2}(r-r_+)(r-r_-),
\end{equation}
with
\begin{equation}
\label{rpm}
\quad r_{\pm}= -\Lambda\pm \sqrt{\Lambda^2+1}.
\end{equation}
Since $r_-<0$ and $0<r_+<1$, it follows that there is a unique fixed point, which is stable, so that $r(t)\rightarrow r_+$ as $t\rightarrow \infty$. Including the resetting condition $r(t)\rightarrow r_0$ then implies that $r(t)\in [r_0,r_+]$ if $r_0<r_+$ and $r(t)\in [r_+,r_0]$ when $r_0>r_+$.
Substituting Eq. (\ref{fL}) into Eq. (\ref{Fir}) with
\begin{align}
\int \frac{ds}{f(s)} &=
\frac{1}{KR_0\sqrt{\Lambda^2+1}}\int \left [ \frac{ ds}{s-r_+}-\frac{ds}{s-r_-}\right ] \\
& =\frac{1}{KR_0\sqrt{\Lambda^2+1}}\left [\ln |r -r_+|-\ln (r-r_-)\right ],\nonumber
\end{align}
we obtain the result
\begin{equation}
F(r)=\frac{KR_0}{2}|r-r_+|^{1-\beta}(r-r_-)^{1+\beta}  
\end{equation}
with 
\begin{equation}
\beta = \frac{ \lambda }{KR_0\sqrt{\Lambda^2+1}}.
\end{equation}
We then have two distinct cases. If $r_0<r_+$ then $A_-=0$ whereas if $r_0>r_+$ then $A_+=0$. Hence, the solution in either case can be written as
\begin{align}
q^*(r;R_0)&=\frac{2\lambda}{KR_0}\left (\left |\frac{r-r_+}{r_0-r_+}\right |\right )^{\beta-1}\left (\frac{r-r_-}{r_0-r_-}\right )^{-1-\beta} \nonumber \\
&\quad \times \frac{1}{|r_+-r_0|(r_0-r_-)},\quad r\in [r_{\min},r_{\max}],
\label{q1}
\end{align}
with $r_{\min}=\min \{r_0,r_+\}$ and $r_{\max}=\max \{r_0,r_+\}$. Note that $q^*$ is parameterized by $R_0$.
Using the identities
\begin{align}
\frac{d}{dr}\left (\frac{r-r_+}{r-r_-}\right )^{\beta}=\frac{\beta (r_+-r_-)}{(r-r_+)(r-r_-)}\left (\frac{r-r_+}{r-r_-}\right )^{\beta}
\end{align}
and
\begin{equation}
\beta (r_+-r_-)=2\sqrt{1+\Lambda^2}\frac{\lambda}{KR_0\sqrt{1+\Lambda^2}}=\frac{2\lambda }{KR_0}
\end{equation}
we see that for $r_0>r_+$
\begin{equation}
q^*(r;R_0)=\left (\frac{r_0-r_-}{r_0-r_+}\right )^{\beta}\frac{d}{dr}\left (\frac{r-r_+}{r-r_-}\right )^{\beta}
\label{qdiff}
\end{equation}
and hence
\begin{equation}
\int_{r_+}^{r_0}q^*(r;R_0)dr=1.
\end{equation}
The same normalization condition holds when $r_0<r_+$.
It also follows from Eq.(\ref{q1}) that $q^*(r)$ is singular at $r=r_+$ when $\beta <1$, which is equivalent to the condition
\begin{equation}
\lambda < \lambda_c =KR_0\sqrt{1+\Lambda^2}.
\end{equation}
 This is illustrated in Fig. \ref{fig7}, which shows example plots of the NESS $q^*(x)$ for various values of $R_0$ and $\lambda$. Note that if $R_0\ll 1$ then the NESS $q^*(r)$ is localized around $0^+$ indicating that the oscillators are in an incoherent state. This is consistent with the behavior shown in Fig. \ref{fig4}.

\begin{figure}[b!]
\centering
\includegraphics[width=10cm]{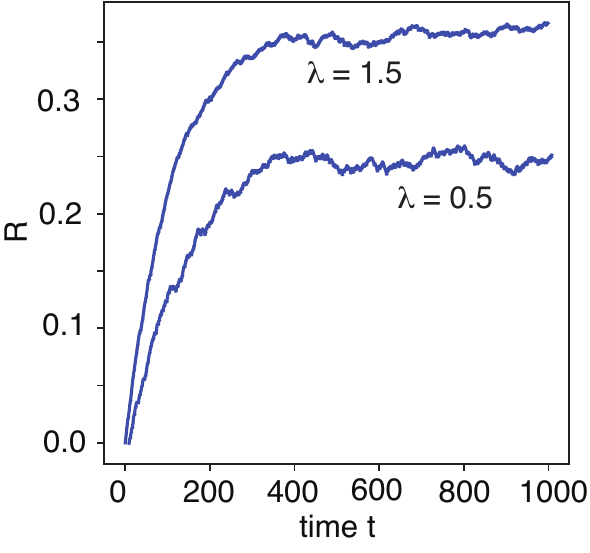} 
\caption{Plot of the slow amplitude $R(t)$ in the low density regime for (a) $\lambda = 0.5$ and (b) $\lambda = 1.5$. Other parameters are $R(0)=0$, $\Delta =1$, $K=1$, $r_0=0.5$ and $\sigma_0 =0.01$.} 
\label{fig8}
\end{figure}

So far we have taken the external amplitude to be fixed at a value $R_0$. However, we can use the NESS for $r$ to obtain an approximate solution for the slow variable $R(t)$ by applying an averaging theorem to Eq. (\ref{OArR2}b) and assuming that the stochastic dynamics of $r(t)$ is ergodic. More specifically, we replace Eq. (\ref{OArR2}b) by the averaged system
\begin{equation}
\frac{dR}{d\tau} = \E[r|R] -R
\label{av}
\end{equation}
where $\tau=\epsilon t$, $\epsilon =\sigma_0K \ll 1$, and
\begin{equation}
\E[r|R] =\lim_{T\rightarrow \infty} \int_0^T r(t)dt =\int_0^1q^*(r,R)rdr.
\end{equation}
Assuming that $R(\tau)\rightarrow R^*$ in the limit $t\rightarrow \infty$, we deduce that if the averaging theorem holds then the asymptote $R^*$ should approximately satisfy the self-consistency condition
\begin{equation}
R^*=\E[r|R^*].
\end{equation}
The conditional expectation can be determined using the solution (\ref{q1}). For the sake of illustration, suppose that $r_0>r_+$. Then
\begin{align}
\E[r|R^*] &=\left (\frac{r_0-r_-}{r_0-r_+}\right )^{\beta} \int_{r_+}^{r_0}r\frac{d}{dr}\left (\frac{r-r_+}{r-r_-}\right )^{\beta}dr \nonumber \\
&=r_0-\left (\frac{r_0-r_-}{r_0-r_+}\right )^{\beta}  \int_{r_+}^{r_0}\left (\frac{r-r_+}{r-r_-}\right )^{\beta}dr,
\label{ave}
\end{align}
after integrating by parts. (Recall from Eq. (\ref{rpm}) that $r_{\pm }$ depend on $R_0=R^*$.)
We can confirm the validity of the slow/fast analysis by numerically solving the planar system (\ref{OArR2}a,b) in the low density regime, as illustrated in Fig. \ref{fig8}. We see that $R^*\approx 0.25$ when $\lambda=0.5$ and $R^*\approx 0.35$ when $\lambda=1.5$. Plugging in the numerical estimates for $R^*$ into Eq. (\ref{ave}) and numerically evaluating the remaining integral we find that $\E[r|R^*] \approx 0.244$ for $\lambda=0.5$ and $\E[r|R^*]. \approx 0.36$ for $\lambda=1.5$. Both of these results agree very well with the corresponding asymptotes $R^*$ shown in Fig. \ref{fig8}.

\subsection{High density regime}

\begin{figure}[t!]
\centering
\includegraphics[width=10cm]{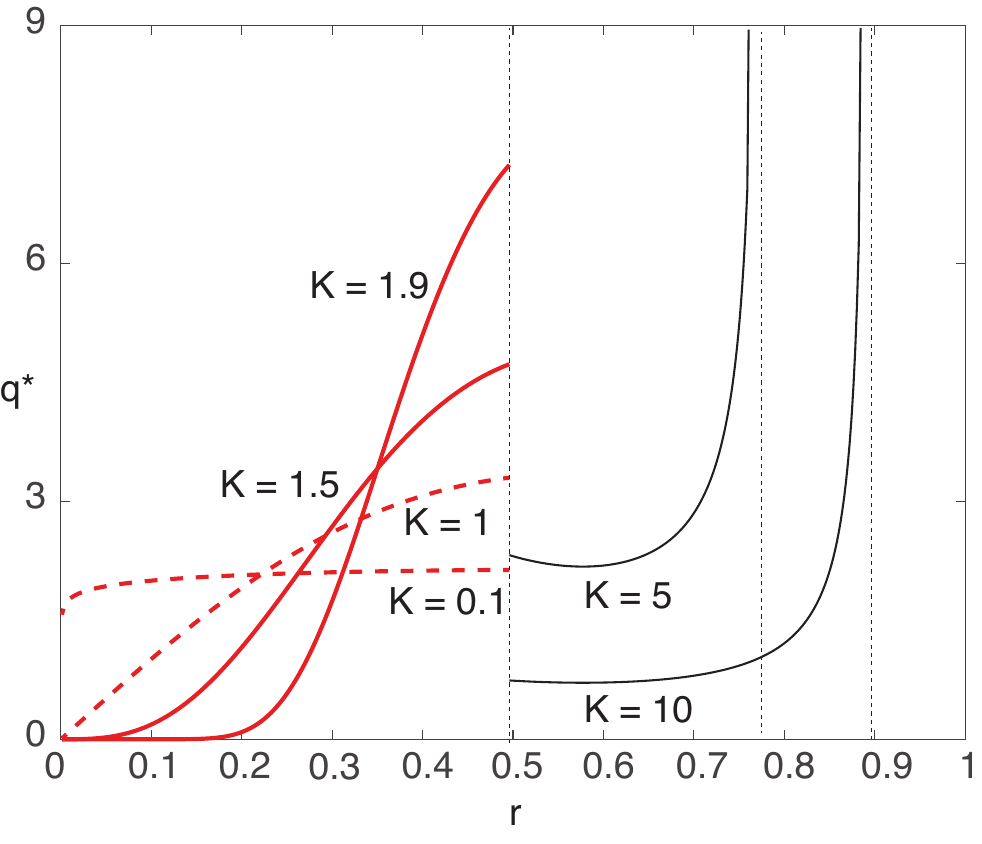} 
\caption{Plots of the NESS $q^*(r)$ given by Eq. (\ref{pR1}) as a function of the order parameter $r$ for various coupling strengths $K$, with $r_0=0.5$,  $\Delta=1$, and $\lambda=1$.}
\label{fig9}
\end{figure}

In the high density regime ($\sigma_0 K\gg \Delta$) the amplitude $R(t)$ is the fast variable and $r(t)$ is  the slow variable. In particular, taking $\sigma_0\rightarrow \infty$, we can set $R(t)=r(t)$ in Eq. (\ref{OArR2}a) to give Eq. (\ref{SHS}) 
with
\begin{equation}
\label{fH}
f(r)=\frac{Kr}{2}(r^2-\Gamma),\quad \Gamma = 1-\frac{2\Delta}{K} .
\end{equation}
Eqs. (\ref{SHS}) and (\ref{fH}) are identical in form to the stochastic dynamics of the order parameter for the classical Kuramoto model with global resetting \cite{Sarkar22,Bressloff24a}. We briefly summarize the results obtained for the NESS on the corresponding OA manifold in order to compare with the low density case. Substituting Eq. (\ref{fH}) into (\ref{Fir}) leads to the following results\cite{Sarkar22}. If $K<K_c=2\Delta$ then
\begin{align} 
q^*(r)&=\frac{2\lambda }{K r^3}\left (1+\frac{|\Gamma|}{r_0^2}\right )^{\alpha}\left (1+\frac{|\Gamma|}{r^2}\right )^{-(1+\alpha)}, \ r\in [0,r_0]\label{pR1}
\end{align}
and $q^*(r)=0$ for $r\notin [0,r_0]$, where
\begin{equation}
 \alpha=\frac{\lambda}{2\Delta -K}.
\end{equation}
On the other hand, if $K>K_c=2\Delta$ then
\begin{align} 
q^*(r)&=\frac{2\lambda}{K r^3}\left (\left |1-\frac{r_c(K)^2}{r_0^2}\right |\right )^{-|\alpha|}\left (1-\frac{r_c(K)^2}{r^2}\right )^{|\alpha|-1} 
 \mbox{ for } r\in  [r_{\min},r_{\max}],
\label{pR2}
\end{align}
and $q^*(r)=0$ for $r\notin  [r_{\min},r_{\max}]$,
with $r_{\min}=\min \{r_0,r_c(K)\}$, $r_{\max}=\max \{r_0,r_c(K)\}$ and $r_c(K)=\sqrt{1-2\Delta/\K}$.

Plots of $q^*(r)$ for various coupling strengths $K$ are shown in Fig. \ref{fig9}. Although a wider range of NESS profiles can be supported by the OA dynamics in the high density limit including nonmonotonic functions, the general behavior is similar to the low density case. In particular, $q^*(r)$ is restricted to a subinterval of $[0,1]$ with $r_0$ at one end, while the other end may develop a singularity.

 \setcounter{equation}{0}
\section{Discussion} In this paper we studied the dynamics of a Kuramoto model with global phase resetting and indirect coupling mediated by an external medium. We first derived a continuum equation for the population density that stochastically resets to the initial distribution of phases, see Eq. (\ref{SPDE2}). We then projected the infinite-dimensional system onto a four-dimensional OA manifold in which only half the variables reset (subsystem resetting), see Eqs. (\ref{OArR}). Finally, we investigated the stochastic dynamics on the OA manifold using a combination of numerical simulations and slow/fast analysis.
Similar to the case without resetting\cite{Schwab12a}, we found that the OA dynamics is sensitive to the cell density parameter $\sigma_0$. When the oscillators and environment are frequency locked ($\omega_0=\Omega$), the dynamics on the OA manifold reduces to the planar system (\ref{OArR2}). The latter interpolates between the classical Kuramoto model with global resetting\cite{Sarkar22,Bressloff24a} at high densities and the slow/fast system analyzed in Sect. V at low densities. On the other hand, when $\omega_0\neq \Omega$, the bistability of the deterministic model persists in the presence of resetting with the additional feature that subsystem resetting can induce a state transition.

\begin{figure}[t!]
\centering
\includegraphics[width=12cm]{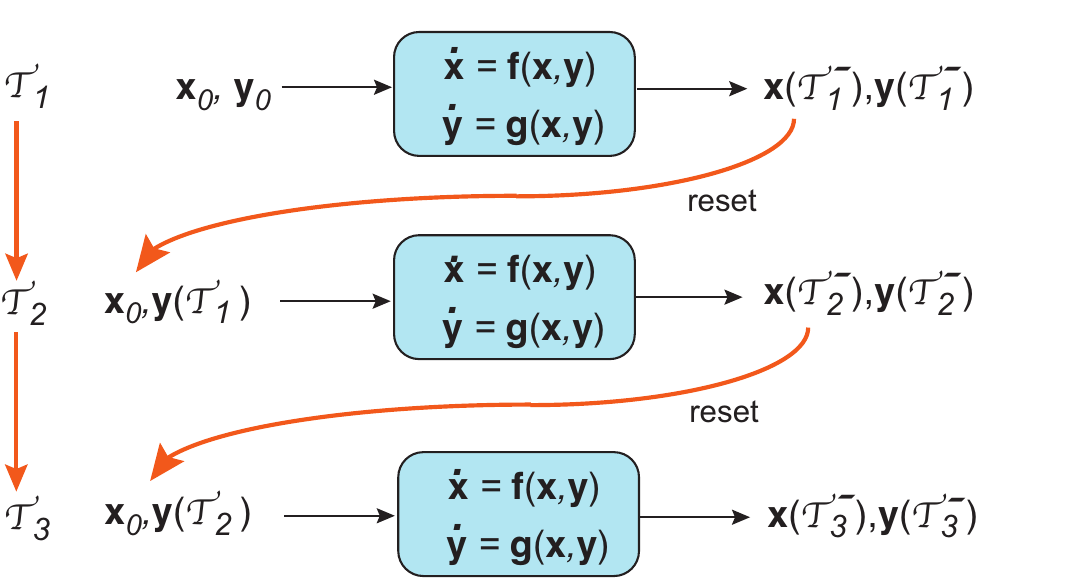} 
\caption{Schematic diagram showing a piecewise deterministic dynamical system with subsystem resetting. Only the first few resetting events are shown.} 
\label{fig10}
\end{figure}

Beyond the particular application to the Kuramoto model, our analysis reveals two general features of piecewise deterministic ODEs with subsystem resetting, see Fig. \ref{fig10},  that warrant further study.
\medskip

\begin{figure}[b!]
\centering
\includegraphics[width=12cm]{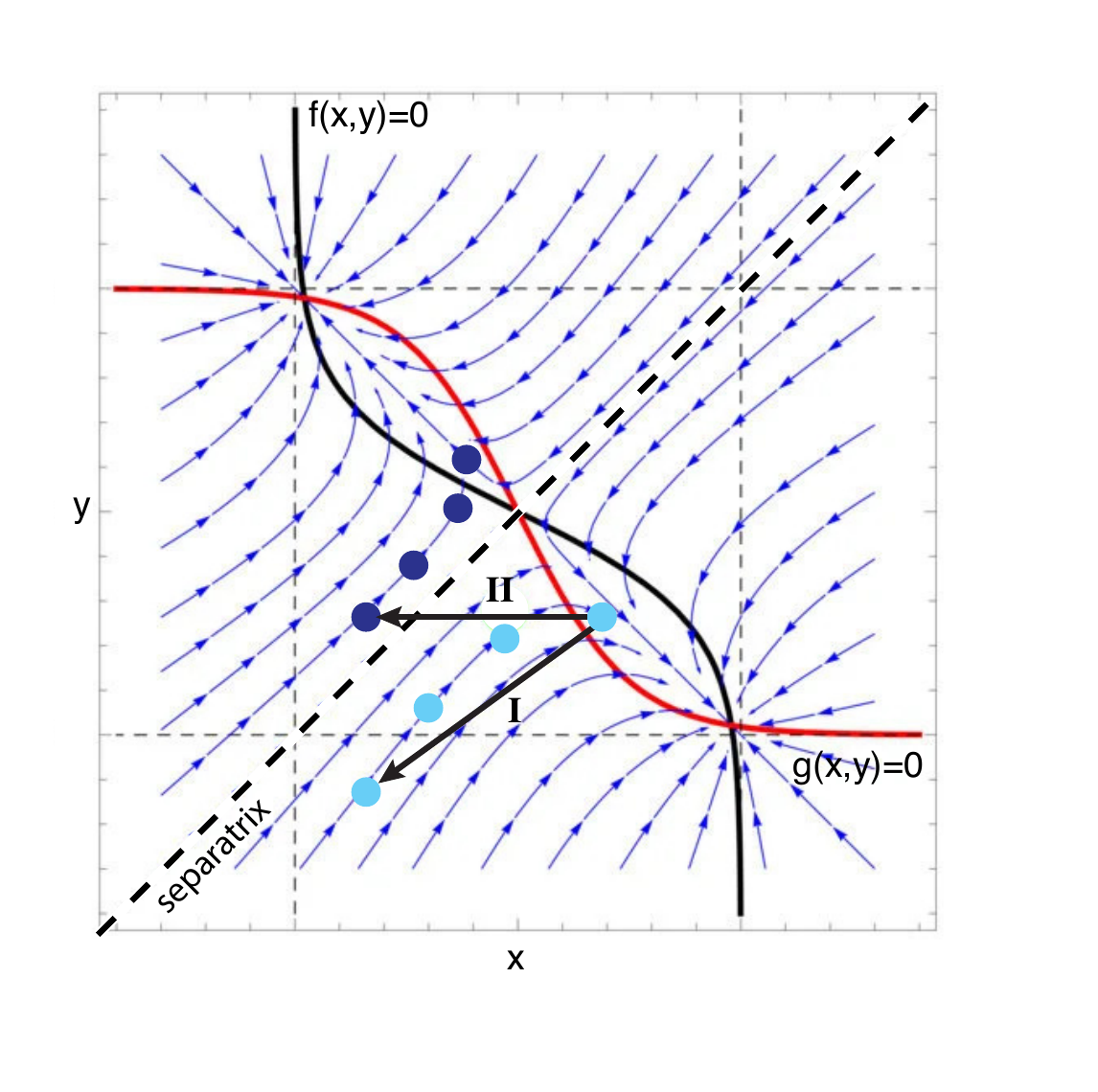} 
\caption{Phase plane diagram for a bistable system subject to resetting. The $x$-nullcline is $f(x,y)=0$ and the $y$-nullcline is $g(x,y)=0$. There are three points of intersection that correspond to a pair of stable fixed points on either side of a saddle. The stable manifold of the saddle (dashed diagonal line) acts as the separatrix between the two basins of attraction. The light filled circles indicate a trajectory that is converging towards the right-hand fixed point prior to reseting to its initial position at $(x_0,y_0)$ -- scenario I. If only the $x$-coordinate resets (subsystem resetting) then the particle jumps to the other basin of trajectory and converges towards the left-hand fixed point prior to the next resetting event -- scenario II.} 
\label{fig11}
\end{figure}

\noindent {\em (i) Multistability}. Suppose that a deterministic system $\dot{\bf x}={\bf f}(\x)$, $\x\in \R^{\calN}$, is multistable. This means that the long-time behavior of the system depends on the basin of attraction of the initial state $\x(0)=\x_0$. Under the total resetting rule $\x(\calT_n^-)\rightarrow \x_0$ for $n\geq 1$, the system remains in the same basin of attraction for all $t>0$ and thus cannot transition to another stable state. On the other hand, if only a subset of variables reset, then noise-induced transitions can occur even though there are no other sources of noise. For the sake of illustration, consider the subsystem resetting rule 
$x_j(t) \rightarrow x_{j,0}$ for $j=1,\ldots, \calM$ with $\calM <\calN$. After each resetting event we have a new random initial condition
$(x_{1,0},\ldots x_{\calM,0},x_{\calM+1}(\calT_n),\ldots x_{\calN}(\calT_n))$. This new initial condition could, in principle, lie in a different basin of attraction thus resulting in a noise-induced transition. This is illustrated schematically in Fig. \ref{fig11} for a bistable planar system with two stable fixed points separated by a saddle. The stable manifold of the saddle (dashed diagonal) acts as the separatrix between the two basins of attraction. Subsystem resetting allows a trajectory to cross the separatrix. However, as we found in the examples shown in Figs. \ref{fig5} and \ref{fig6}, this transition is irreversible. Fig. \ref{fig11} does suggest one way to reverse the transition, namely, using different reset positions in the two basins of attraction. However, this is rather contrived. In future work it would be interesting to explore in more detail the role of subsystem resetting in noise-induced transitions. What is the interplay between subsystem resetting and other noise sources? Can subsystem resetting induce a cascade of state transitions in the case of multiple fixed points? Is it possible to estimate the expected time of a state transition?

\medskip

\noindent {\em (ii) Slow/fast analysis}. Consider a slow/fast dynamical system of the form
\begin{subequations}
\label{sf}
\begin{align}
\frac{d\x}{dt}&={\bf f}(\x,\y),\\ \frac{d\y}{dt}&=\epsilon {\bf g}(\x,\y)
\end{align}
\end{subequations}
with $0<\epsilon \ll 1$. Suppose that only the fast variables reset, $\x(\calT_n^-)\rightarrow \x_0$. In the limit $\epsilon \rightarrow 0$ we can treat the slow variables as fixed and set $\y(y)=\y_0$ in Eq. (\ref{sf}a). We thus have the following reduced system with resetting:
\begin{align}
\frac{d\x}{dt}&={\bf f}(\x,\y_0)+h(t)[\x(t^-)-\x_0].
\end{align}
with $\y_0$ treated as a set of parameters.
Let $p(\x,t;\y_0)$ denote the corresponding parameterized probability density, which evolves according to
\begin{equation}
\frac{\partial p}{\partial t}=-{\bm \nabla}_{\x}\cdot [{\bf f}p]-rp+r\delta(\x-\x_0).
\end{equation}
Suppose that there exists a unique NESS $p^*(\x;\y_0)$ which is ergodic. We then define the averaged equation
\begin{align}
\frac{d\y}{d\tau}&={\bf g}(\langle \x(\tau)\rangle,\y(\tau)),\quad \langle \x(\tau)\rangle =\int \x p^*(\x,\y(\tau))d\x,
\end{align}
with $\tau=\epsilon t$. For a specific model it would be necessary to determine how well the solution of the averaged system approximates the solution of the full system (\ref{sf}) over a given time interval.

\end{document}